\documentclass[12pt]{iopart}

\usepackage{iopams}
\usepackage{amsopn} 
\usepackage{graphicx}
\usepackage{subfigure}
\usepackage{rotating}
\usepackage{mathrsfs} 
\usepackage{array}
\newcolumntype{L}{>{\displaystyle}l}
\newcolumntype{R}{>{\displaystyle}r}
\newcolumntype{C}{>{\displaystyle}c}

\usepackage{xcolor}
\usepackage{hyperref}

\newcommand{\vc}[1]{{\bi #1}}
\def\R{\mathbb{R}}
\def\const{\mathrm{const}}
\def\scrK{\mathcal{K}}
\def\gI{\hat g}
\def\MI{\hat M}
\def\gS{g^{\mathrm{S}}_{M}}
\def\gSc{g^{\mathrm{S}}_{M,\vc c}}
\def\gBL{g^{\mathrm{BL}}}
\def\gBrill{g_{\mathrm{Brill}}}
\def\psiS{\psi^{\mathrm{S}}_M}
\def\psiBL{\psi^{\mathrm{BL}}}
\def\qdn{q_{dn}}
\def\qnd{q_{nd}}
\def\a{\alpha}
\def\b{\beta}
\def\d{\delta}
\def\ta{\theta}
\newcommand{\Scri}{\mathscr{I}}
\DeclareMathOperator{\sech}{sech}
\newcommand{\norm}[1]{\Vert #1 \Vert}
\newcommand{\eqref}[1]{\eref{#1}}
\newcommand{\text}[1]{\mbox{#1}} 

\newcommand{\coloneqq}{\mathrel{%
    \mathrel{%
        \rlap{\raisebox{0.35ex}{$\vcenter{\hbox{.}}$}}%
        \raisebox{-0.35ex}{$\vcenter{\hbox{.}}$}%
    }=%
}}

\newenvironment{stretchedarray}[2][1.2]{%
    \settowidth{\arraycolsep}{$\mskip2.5mu$}
    \renewcommand{\arraystretch}{#1}%
    \begin{array}{@{}#2@{}}%
}{%
    \end{array}%
}

\newenvironment{subeqns}[1][]{%
    \ifx#1\\\\\else
        \addtocounter{equation}{1}%
        \unskip%
        \raisebox{0pt}[0pt][0pt]{
            \begin{minipage}{0cm}%
                \eqnarray\nonumber\endeqnarray%
                #1
            \end{minipage}%
        }%
        \addtocounter{equation}{-1}%
    \fi%
    \numparts%
}{%
    \endnumparts%
    \ignorespacesafterend%
}

\begin{document}

\title{Numerical Approach for Corvino-Type Gluing of
       Brill-Lindquist Initial Data}

\author{Daniel Pook-Kolb}

\address{Max Planck Institute for Gravitational Physics 
(Albert Einstein Institute), Leibniz Universit\"at Hannover, 
Callinstrasse 38, D-30167 Hannover, Germany}
\ead{daniel.pook.kolb@aei.mpg.de}

\author{Domenico Giulini}
\address{Institute for Theoretical Physics, Leibniz Universit\"at Hannover,
         Appelstrasse 2, D-30167 Hannover, Germany}
\vspace{5pt}
\address{Center of Applied Space Technology and Microgravity,\\
         University of Bremen, Am Fallturm 1, D-28359 Bremen, Germany}
\ead{domenico.giulini@itp.uni-hannover.de}

\begin{abstract}
    Building on the work of Giulini and Holzegel (2005), a new numerical
    approach is developed for computing Cauchy data for Einstein's
    equations by {\em gluing} a Schwarzschild end to a Brill-Lindquist metric
    via a Corvino-type construction.
    In contrast to, and in extension of, the numerical strategy of Doulis and
    Rinne (2016), the overdetermined Poisson problem resulting from the Brill
    wave ansatz is decomposed to obtain two uniquely solvable problems. A
    pseudospectral method and Newton-Krylov root finder are utilized to
    perform the gluing.
    The convergence analysis strongly indicates that the numerical strategy
    developed here is able to produce highly accurate results. It is observed
    that Schwarzschild ends of various ADM masses can be glued to the same
    interior configuration using the same gluing radius.
\end{abstract}

\vspace{2pc}
\noindent{\it Keywords}:  General Relativity, numerical relativity,
initial-value problem



\section{Introduction}
The theoretical possibility of gravitational-wave 
generation through the merger of two black holes 
has recently received spectacular confirmation through 
several earth-bound detections of gravitational waves 
by the 
LIGO-Virgo-collaboration~\cite{Abbott:2016,Abbott:2017a,Abbott:2017b}. 
Also, these observations were most recently complemented 
by the no less spectacular detection of the gravitational-wave 
signal from the merger of two neutron-stars~\cite{Abbott:2017c}. 
From a theoretical point of view the problem of modelling 
binary neutron-star systems differs drastically from the 
former insofar as it involves a complicated coupled system of 
differential equations that governs both, the dynamics of the 
gravitational field and that of the neutron matter composing the 
stars. In contrast, the theoretical analysis of the dynamics of 
black holes ``only'' involves the sourceless (i.e. vacuum) 
Einstein equations of General Relativity. Hence, in terms of 
differential equations, black holes poses the much ``cleaner'' 
problem. 

But even in this ``simpler'' case there are still 
unresolved mathematical issues of undeniable physical 
relevance in connection with the initial-value formulation 
of the matter-free field equations of General Relativity. 
We recall that these are 10 coupled non-linear 
but quasi-linear partial differential equations of second 
order for the ten components $g_{\mu\nu}$ of the space-time 
metric. Here, as usual, greek indices range in the set 
$\{0,1,2,3\}$ and for definiteness we state that we use the 
``mostly-plus'' convention concerning the signature, though 
this will be irrelevant for the rest of this paper. This 
system of differential equations can be cast into the form 
of a Hamiltonian system 
(see, e.g., \cite{Giulini:SpringerHandbookSpacetime}) 
which reveals its hybrid-nature in the following sense. 
Six out of the ten equations are of (underdetermined) 
hyperbolic type and hence comprise the evolutionary content 
of Einstein's equations, whereas the remaining four equations 
are of (underdetermined) elliptic type. The latter are 
constraints in the sense that they only involve the initial 
data and their spatial derivatives, but not their time 
derivatives. Hence they properly constrain the initial data 
themselves, rather than their evolution. In this paper we will 
be exclusively concerned with these constraints. 

There exist a considerably large body of knowledge concerning 
explicit analytic expressions of solutions to the constraints 
representing initial data for two or more black holes. 
A classic paper is \cite{brill1963interaction}, in which 
initial data are given for two black holes without orbital and 
spin angular momentum at the moment of rest. For such 
``time symmetric initial data'' (vanishing linear and angular 
momenta), and under the assumption that the Cauchy hypersurface 
is conformally flat, the constraints simply reduce to a single 
Laplace equation for the conformal factor; see, e.g., \cite{Giulini:SpringerHandbookSpacetime} for a review. Hence a
whole arsenal of techniques, like Thomson's method of images 
familiar from electrostatics, can be employed to construct 
solutions with various kinds of symmetries \cite{Giulini:1998a},
even in the time-asymmetric case; e.g., \cite{Bowen.York:1980}
\cite{Brandt.Bruegmann:1997}. 

\begin{figure}
\label{fig:JunkRadiation}
\centering
\includegraphics[scale=1.2]{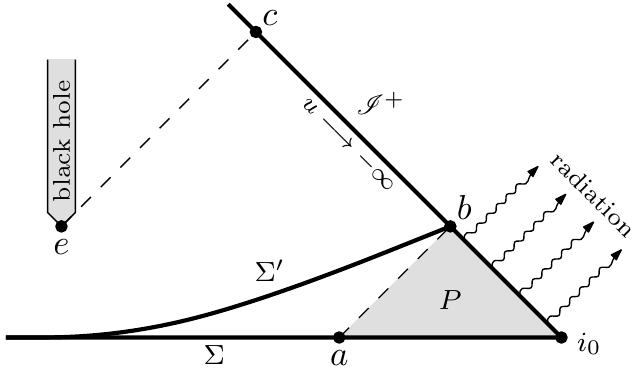}
\caption{\label{fig:radiation}
Conformal (Penrose) diagram of spacetime with two spacelike 
hypersurfaces $\Sigma$ and $\Sigma'$ ending at spacelike 
infinity $i_0$ and future lightlike infinity $\Scri^+$,
respectively. $\Sigma$ depicts an asymptotically flat Cauchy 
surface, whereas $\Sigma'$ is asymptotically hyperboloidal 
and not Cauchy. $u$ denotes the standard Bondi parameter 
along the null generators of $\Scri^+$. The difference between 
the ADM-mass of $\Sigma$ (computed at $i_0$) and the Bondi 
mass of $\Sigma'$ (computed at the intersection 2-sphere 
between $\Sigma'$ and $\Scri^+$, here denoted by the point 
$b$) must be due to gravitational radiation escaping between 
$i_0$ and $b$. This radiation originates from the causal past 
$P$ of the region $\overline{i_0b}\subset\Scri^+$, whose 
intersection with $\Sigma$ is the region $\overline{ai_0}$. 
In this sense we say that the data on $\Sigma$ contain 
radiation in that region. Any gravitational radiation emerging 
from an (quasi localized) event $e$, e.g. the formation of a 
black hole due to binary mergers, cannot reach $\Scri^+$ 
before the intersection $c$ of the future light-cone at 
$e$ with $\Scri^+$. In that sense, any radiation reaching 
$\Scri^+$ before $c$, like that explicitly shown in the 
figure, is considered ``junk''.}
\end{figure}

However, there is one physical issue that affects all these 
data alike, even the most simple ones, which usually goes 
under the name of ``junk radiation''. That is to say, all 
these data already contain initial gravitational radiation 
that fills space up to infinity; 
compare figure\,\ref{fig:radiation}. That radiation is clearly 
seen in numerical simulations and needs to be distinguished 
and separated from that radiation that is dynamically
produced by the process of black-hole collision. Usually 
this is pragmatically done by just not counting that radiation
which identifies itself as being ``junk'', i.e. \emph{not}  
produced in the scattering process, by arriving too early
at large spatial distances, i.e. outside the causal future of
the merging event. But it is clear that this pragmatic procedure,
albeit efficient, is not entirely satisfying, certainly not from 
a theoretical and conceptual point of view. Hence the question 
arises of how to modify the known multi-black-hole initial data 
sets, so as to remove the junk radiation while keeping the other 
aspects intact, like the presence of black holes, their masses,
mutual distances, and possible other parameters.

Analytically, the existence of initial gravitational radiation 
that spreads all the way to infinity has been shown 
\cite{Valiente-Kroon:2003} to relate to the Newman-Penrose 
constants \cite{Newman.Penrose:1965} in case of time-symmetric 
and conformally flat initial data sets \cite{Valiente-Kroon:2003}. 
Furthermore, it has been shown by the same author in 
\cite{Valiente-Kroon:2004} that non-vanishing Newman-Penrose 
constants impose obstructions to smoothness in the transition 
from spatial infinity ($i_0$) to null infinity $(\Scri^\pm)$. 
It has even been conjectured in \cite{Valiente-Kroon:2004}
that such a smooth transition requires the (time symmetric and 
conformally flat) data to be \emph{exactly} Schwarzschild in a 
neighbourhood of $i_0$, though recent work \cite{Hintz.Vasy:2017}
on spacetimes admitting polyhomogeneous expansions of the metric 
at infinity (spacelike and null) suggests that the conjecture is 
false (too strong).\footnote{We thank Piotr Chru\'sciel for pointing 
this out.}

For us the desired modification of data would consist in 
removing all radiation outside some sphere and to maintain 
the old data set inside some smaller sphere still containing 
all the black holes. Their masses, distances etc.~would then 
be preserved. Clearly, a sufficient condition (possibly also
necessary, if the conjecture mentioned above should indeed
be true) for really removing the radiation and obtain a 
smooth future null infinity $\Scri^+$ (at least in a 
neighbourhood of $i_0$) is to modify the data outside the 
larger sphere so as to become \emph{exactly} Schwarzschild 
(in the time symmetric case). 

A priori it is not at all obvious that such a modification 
exists at all, as the data have to satisfy the mentioned 
constraints which are of (underdetermined) elliptic type. 
Fortunately, such a modification is indeed possible, as was 
shown in 2000 through a non-constructive existence proof 
\cite{corvino2000}. The pattern of modification is precisely 
the one envisaged above, where outside a larger sphere one 
obtains the exterior Schwarzschild data, inside the inner 
sphere the restriction of the original data set, and in the 
annular region inbetween the two spheres one has some interpolating 
metric that furnishes a smooth transition between the two. 
The proof in \cite{corvino2000} merely asserts the existence 
of a transition region without further control of its size, 
and the existence of a transition metric without further control
of its properties. 

An attempt to make this procedure explicit in the simplest 
case of the two-hole Brill-Lindquist data was made in 
\cite{giulini2005corvino}. Since these initial data are 
axisymmetric, the central idea here was to also restrict 
the transition metric to a simple axisymmetric form of 
Brill waves \cite{brill1959positive}. A central concern
of  \cite{giulini2005corvino} was to find out whether this 
Corvino-like modification of the two-hole Brill-Lindquist 
data set using Bill waves could be used to reduce the overall 
ADM energy. This is not obvious since although one clearly 
removed all gravitational radiation outside the larger 
sphere, one possibly also re-introduces new one in the 
transition region, possibly by an inappropriate choice
of the transition metric. On the other hand, one suspects 
on physical grounds that such a reduction of energy should 
clearly be possible, given the existence of junk radiation 
in the first place. Presumably one just has to make the 
right use of the large freedom in choosing the transition 
metric to achieve that end, but a clear proof that this is
indeed possible was, and still is, missing. In this paper 
we contribute towards a decision on that question.

Building on ideas of \cite{giulini2005corvino} some recent 
numerical implementation 
\cite{rinne2015numerical,rinne2016numerical} found, 
somewhat surprisingly, no evidence that an overall 
mass reduction is indeed possible, though the converse 
was also not ruled out and a question remained open. As the
authors themselves admitted, their approach severely 
restricted the set of transition metrics, thereby possibly 
missing out the ``good ones''. Here we present a novel 
approach to the numerical implementation that allows 
to avoid the overdetermined nature of the boundary-value 
problems encountered in \cite{rinne2015numerical,rinne2016numerical}. 
This will allow a larger flexibility in choosing 
transition metrics, possibly including energy reducing 
ones. But that possibility still awaits being turned into reality.

\section{The Gluing Construction}

We will start with a short summary of the main statement in
\cite{corvino2000}.
For any smooth, asymptotically flat, and scalar flat metric 
$\gI$ on $\R^n$, $n \geq 3$, which is conformally
flat at infinity and has positive mass $\MI$, and any 
compact set $\scrK \subset \R^n$, there is a scalar flat 
metric $g$ on $\R^n$ which is exactly Schwarzschild
near infinity yet satisfies $g \equiv \gI$ inside $\scrK$.
Here, ``near infinity'' means outside a compact set.
In the proof, a candidate metric $\tilde g$ is constructed 
which simply blends smoothly between the interior metric 
$\gI$ and a Schwarzschild end $\gSc$ of ADM mass $M$ and 
center of mass $\vc c$. This blending is confined to an 
{\em annular region} $A_R \coloneqq B_{2R} \setminus \overline B_R$,
where $R$ is the {\em gluing radius} and
$B_R$ is an open ball of radius $R$ (with respect to the 
asymptotically flat coordinates). The candidate metric 
$\tilde g$ will then in general not be scalar flat,
i.\,e.~we have $R(\tilde g) \neq 0$ on $A_R$.
Corvino now proves that for sufficiently large $R$
there exists a mass $M$, center of mass $\vc c$,
and a smooth deformation $h$ with support on $A_R$
such that $R(\tilde g + h) \equiv 0$.
Due to the localized nature of the gluing, the 
proof works for any conformally flat end of a scalar 
flat asymptotically flat metric $\gI$, which 
includes non-trivial black hole data.
However, the proof does not indicate how the deformation 
$h$ or the parameters $M$ and $\vc c$ of the 
Schwarzschild end can be obtained in practice.

In order to implement such a construction in a concrete case,
Giulini and Holzegel \cite{giulini2005corvino} glue a 
Schwarzschild end, i.\,e.~a spacelike $t=\const$ slice 
of an exterior Schwarzschild metric in isotropic coordinates,
\begin{equation}\label{eq:gS}
    \gS = (\psiS)^4\ \d
                \coloneqq \left( 1 + \frac{M}{2\norm{\vc x}} \right)^4\ \d,
\end{equation}
to a Brill-Lindquist interior metric
\begin{equation}\label{eq:gBL}
    \gBL = (\psiBL)^4\ \d
                \coloneqq \left(
                    1
                    + \frac{m}{2\norm{\vc x - \vc x_0}}
                    + \frac{m}{2\norm{\vc x + \vc x_0}}
                \right)^4\ \d,
\end{equation}
where $\vc x_0 = (0, 0, d/2)^T$ and $\d$ is the flat 3-metric.
Equation~\eqref{eq:gBL} describes two black holes of equal 
mass at the moment of time symmetry with positions on the 
$z$-axis symmetrically about the coordinate origin.
Note that \eqref{eq:gS} and \eqref{eq:gBL} have ADM masses
$M$ and $2m$, respectively, and naturally $\vc c = 0$.

Since both $\gS$ and $\gBL$ have cylindrical symmetry, the 
initial blending can be performed using a conformally 
transformed {\em Brill wave} \cite{brill1959positive}, 
which is the most general axisymmetric 3-metric. In 
spherical polar coordinates it reads
\begin{equation}\label{eq:brill_wave}
    \gBrill = \psi^4 \left(
        e^{2q} \left( dr^2 + r^2\,d\ta^2 \right) + r^2 \sin^2\ta\ d\phi^2
    \right).
\end{equation}
Here, $\psi$ and $q$ are functions of $r$ and $\ta$
satisfying the following conditions
stated in \cite{brill1959positive}:
\begin{subeqns}[\label{eq:q_conditions}]
    \begin{eqnarray}
        \label{eq:q_conditions_a}
        q &= 0
            \qquad& \text{for $\ta = 0$ and $\ta = \pi$ (on the $z$-axis)},
        \\
        \label{eq:q_conditions_b}
        \frac{\partial q}{\partial \ta} &= 0
            \qquad& \text{for $\ta = 0$ and $\ta = \pi$ (on the $z$-axis)},
        \\
        \label{eq:q_conditions_c}
        q &\in \Or(r^{-2})
            \qquad& \text{for $r \to \infty$},
    \end{eqnarray}
\end{subeqns}
and
\begin{subeqns}[\label{eq:psi_conditions}]
    \begin{eqnarray}
        \label{eq:psi_conditions_a}
        \psi &> 0
            \qquad& \text{everywhere},
        \\
        \label{eq:psi_conditions_b}
        \psi - 1 &\in \Or(r^{-1})
            \qquad& \text{for $r \to \infty$},
        \\
        \label{eq:psi_conditions_c}
        \frac{\partial \psi}{\partial \ta} &= 0
            \qquad& \text{for $\ta = 0$ and $\ta = \pi$ (on the $z$-axis)}.
    \end{eqnarray}
\end{subeqns}
One possible choice for a conformal factor $\psi$ realizing a 
smooth blending between $\gBL$ and $\gS$ would be
\begin{equation}\label{eq:psi_blend}
    \psi(r,\ta) = \b(r,\ta)\; \psiBL(r,\ta) + (1-\b(r,\ta))\; \psiS(r),
\end{equation}
where $\b$ is any smooth cutoff function in $A_R$ satisfying
\begin{subeqns}[\label{eq:beta_conditions}]
    \begin{eqnarray}
        \label{eq:beta_conditions_a}
        \b(r,\ta) &= \cases{1 & for $r \leq R$\\
                            0 & for $r \geq 2R$,}
        \\
        \label{eq:beta_conditions_b}
        \frac{\partial^n \b}{\partial r^n} &= 0
            \quad\text{for all $n \geq 1$ and $r = R$ or $r = 2R$},
        \\
        \label{eq:beta_conditions_c}
        \frac{\partial \b}{\partial \ta} &= 0
            \quad\text{for $\ta = 0$ and $\ta = \pi$}.
    \end{eqnarray}
\end{subeqns}
The first condition ensures that we blend from the Brill-Lindquist 
to the Schwarzschild conformal factor inside $A_R$, while the 
second condition guarantees smoothness of the radial blending.
The third condition follows from \eqref{eq:psi_conditions_c}.

Note that a choice of the form \eqref{eq:psi_blend} and $q \equiv 0$ just
corresponds to a candidate metric $\tilde g$ for which, in general,
$R(\tilde g) \neq 0$ in $A_R$.
However, if we assume the final glued metric $g = \tilde g + h$ to preserve
the cylindrical symmetry, then it can also be expressed in the form of
\eqref{eq:brill_wave}, i.\,e.~$\tilde g + h = \gBrill$ for some choice of
$\psi$ and $q$.
Inserting now \eqref{eq:brill_wave} into $R(\gBrill) = 0$, we get
\begin{equation}\label{eq:poisson_q_full}
    \eqalign{
        \left(
            \frac{\partial^2}{\partial r^2}
            + \frac{1}{r} \frac{\partial}{\partial r}
            + \frac{1}{r^2} \frac{\partial^2}{\partial\ta^2}
        \right) q(r,\ta)
        \cr\qquad
        = -4 \psi^{-1} \left(
            \frac{\partial^2}{\partial r^2}
            + \frac{1}{r} \frac{\partial^2}{\partial\ta^2}
            + \frac{2}{r} \frac{\partial}{\partial r}
            + \frac{\cot\ta}{r^2} \frac{\partial}{\partial\ta}
        \right) \psi(r,\ta),
    }
\end{equation}
which can be read as a 2-dimensional Poisson equation for $q$
\begin{equation}\label{eq:poisson_q}
    \Delta^{(2)} q = f \coloneqq -4 \frac{\Delta^{(3)}\psi}{\psi}.
\end{equation}
If equation~\eqref{eq:poisson_q} is solved with suitable conditions on $q$ and
$\psi$, the scalar curvature vanishes everywhere and the gluing has been
performed.

\subsection{Boundary Conditions}
\label{sub:boundary_conditions}

\begin{figure}[htbp]
    \centering
    \includegraphics{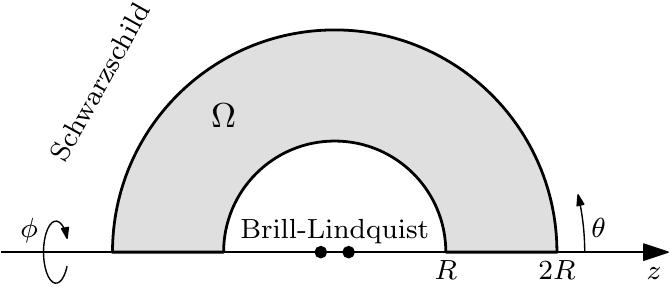}
    \caption{%
        Half annular plane $\Omega$ on which to solve
        equation~\eqref{eq:poisson_q}.%
    }
    \label{fig:domain}
\end{figure}

We require $\gBrill$ to be equal to $\gS$ for $r \geq 2R$ and $\gBL$
for $r \leq R$. This means that $q = 0$ and $\Delta^{(3)}\psi = 0$ outside
$A_R$, thereby trivially satisfying \eqref{eq:poisson_q}.
Hence, equation~\eqref{eq:poisson_q} is to be solved on $A_R$ only,
which, due to the cylindrical symmetry, reduces to the half annular plane
$\Omega$ depicted in figure~\ref{fig:domain}.
Smoothness of $\gBrill$ then implies
\begin{subeqns}[\label{eq:q_radial_conditions}]
    \begin{eqnarray}
        \label{eq:q_radial_conditions_a}
        q &= 0
            \qquad& \text{for $r = R$ and $r = 2R$ (on the radial boundary)},
        \\
        \label{eq:q_radial_conditions_b}
        \frac{\partial^n q}{\partial r^n} &= 0
            \qquad& \text{for all $n \geq 1$, $r = R$ and $r = 2R$}.
    \end{eqnarray}
\end{subeqns}
For the same reasons, conditions \eqref{eq:q_radial_conditions} also apply to
$\Delta^{(3)}\psi$ and hence to $f$.
Fortunately, this simplifies the conditions for $q$:
If \eqref{eq:q_radial_conditions} holds for $n=1$, then by
equation~\eqref{eq:poisson_q} it holds for any $n > 1$.
Combining this with \eqref{eq:q_conditions}, the complete boundary
conditions for $q$ consist of homogeneous Dirichlet and homogeneous Neumann
conditions on $\partial\Omega$.
Clearly, this makes equation~\eqref{eq:poisson_q} overdetermined since the
pure Dirichlet problem on $\Omega$ already has a unique solution,
see Theorem~4.3 in \cite{gilbarg2001elliptic}.
Therefore, the task is now to find a conformal factor $\psi$ blending smoothly
from $\psiBL$ to $\psiS$ such that \eqref{eq:poisson_q} has a solution $q$
satisfying Dirichlet and Neumann conditions.
In \cite{giulini2005corvino} this is called the DN-problem and a solution $q$
a DN-solution.

\section{The Numerical Strategy}

An analytical treatment of the DN-problem has been carried out in
\cite{giulini2005corvino} where an approximation result could be obtained.
The objective of the present paper is to present a strategy for numerically
approximating DN-solutions and find explicit values of the glued Schwarzschild
mass $M$.

\subsection{Previous Results}

The first step towards a numerical solution has been made by Doulis and Rinne
in \cite{rinne2015numerical} and \cite{rinne2016numerical}
and the strategy here extends and corrects their results.
Specifically, Doulis and Rinne employ a numerical method to find a
mass $M$ and a gluing function $\b$ based on iteratively solving problems that
are still overdetermined, i.\,e.~in each step they impose a Neumann condition on
$\partial\Omega$ and additionally a Dirichlet condition on the {\em arches} at
$r=R$ and $r=2R$.
With these boundary conditions, the Poisson equation \eqref{eq:poisson_q} has
no solution on $\Omega$ in general; see section~\ref{sub:boundary_conditions}.

To determine whether it is still possible that the numerical search converges
to an existing solution, Doulis and Rinne carry out convergence tests on their
results.
Since the solutions are computed using a pseudospectral method, the expected
behaviour is exponential convergence of the numerical solutions to the exact
solution for increasing resolution.
An approximation of the error of a particular numerical solution is given by
the last coefficient of the expansion of the solution into the chosen basis
functions \cite{boyd2001chebyshev}.
This argument applies to one-dimensional problems. For higher dimensions, one
typically chooses a particular basis set for each dimension and expands the
solution into the tensor product basis.
The coefficients of solutions of two-dimensional problems then become a matrix
$(a_{kl})$, $k = 0,\ldots,K$, $l=0,\ldots,L$,
where $K$ and $L$ define the resolution, i.\,e.~the number of basis
functions to consider.
The question now is how to approximate the error using the
``last coefficient'' in case of a coefficient matrix.
In \cite{rinne2016numerical}, Doulis and Rinne choose $a_{KL}$ and plot how it
decays exponentially in $K$ when increasing $K$ and $L$ at the same time.
They also plot how the $L^2$-norm of the difference between lower resolution
solutions and a reference solution of high resolution decays.

\begin{figure}[htbp]
    \centering
    \includegraphics[trim=50 18 10 28,clip,width=0.33\linewidth]{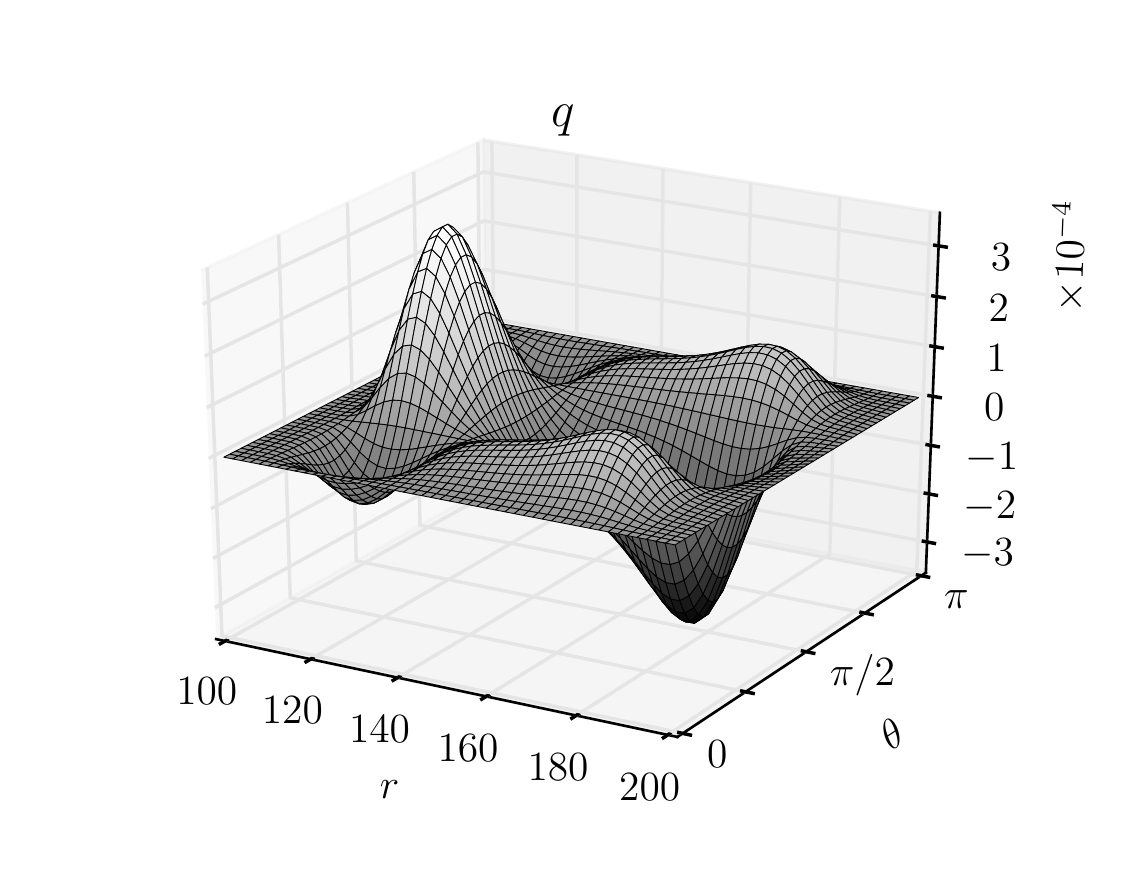}
    \caption{%
        Result of our implementation of the strategy of Doulis and Rinne
        corresponding to the case of figure~6(b) in \cite{rinne2016numerical}
        (i.\,e.~$m=2$, $d=10$, $R=100$).
        Shown is the numerical solution $q$ of the Poisson problem
        \eqref{eq:poisson_q} after the gluing function $\b(r,\ta)$ had been
        determined as in \cite{rinne2016numerical} at the same numerical
        resolution of $K=L=25$. The found ADM mass is
        $M = 4.000\;027\;17$.
    }
    \label{fig:our_doulis_rinne_q}
\end{figure}

\begin{figure}[htbp]
    \centering
    \subfigure[]{
        \includegraphics[trim=2 10 10 0,clip,width=.5\textwidth]{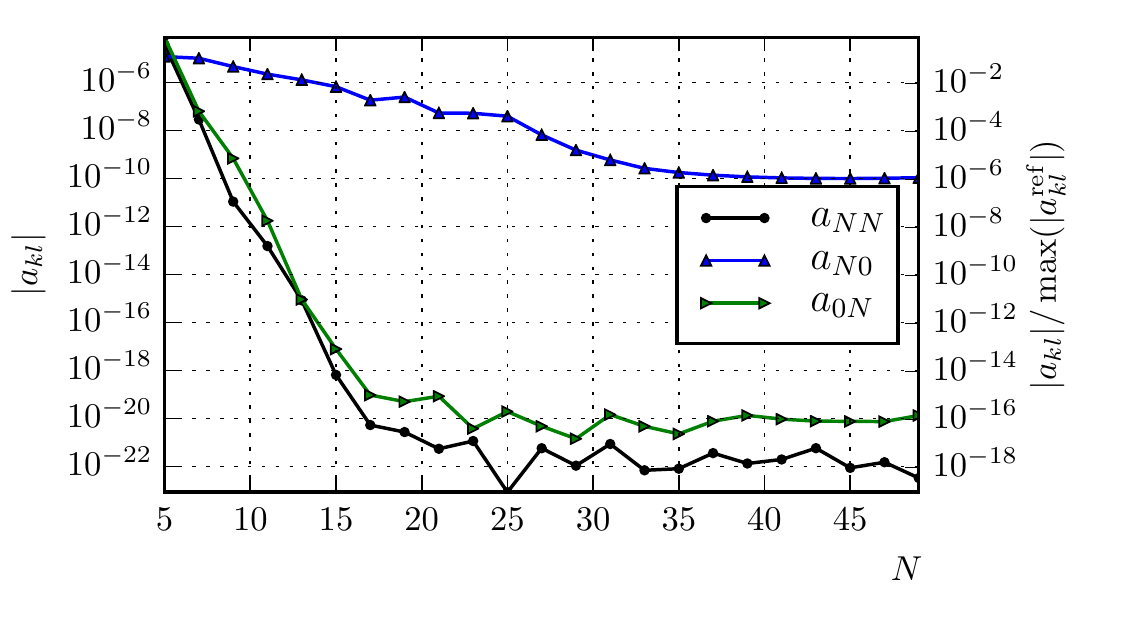}
        \label{fig:coeff_convergence_aNN}
    }%
    \subfigure[]{
        \includegraphics[trim=0 10 14 0,clip,width=.5\textwidth]{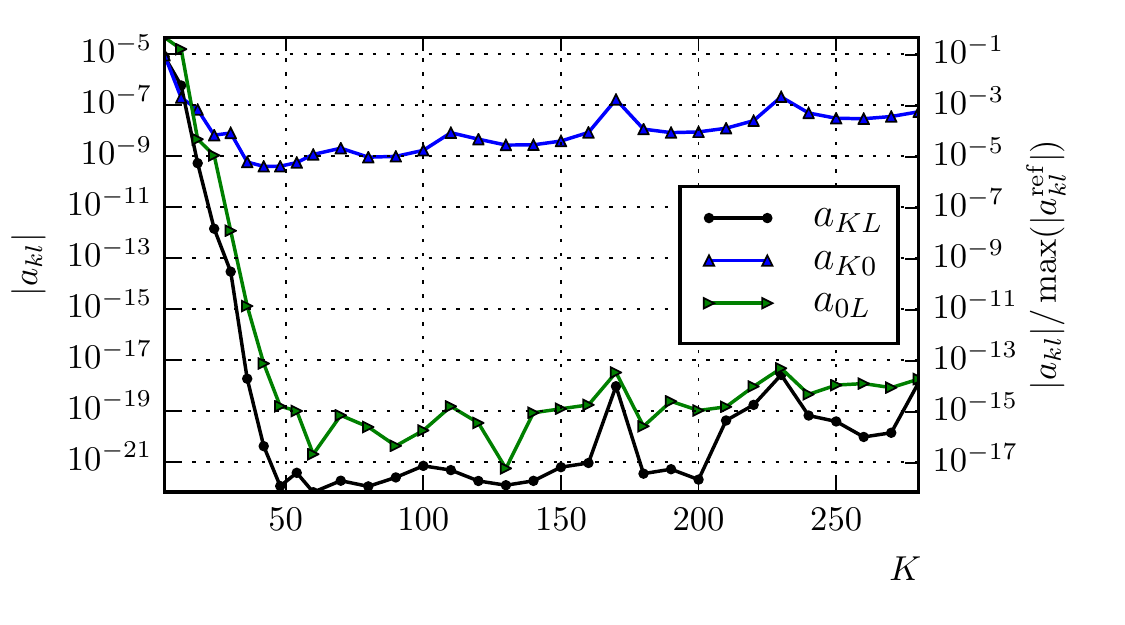}
        \label{fig:coeff_convergence_aKL}
    }
    \caption{%
        Convergence of three different choices of a ``last coefficient'' for
        numerical resolutions of (a)~$K=L=:N$ and
        (b)~$L=\lfloor K/3 \rfloor$ (i.\,e.~the integer part of $K/3$).
        As explained in section~\ref{sub:numerical_setup}, the
        $\pi/2$-symmetry effectively removes half of the coefficients in the
        angular ``direction'', leaving in the end $\lfloor K/2 \rfloor$
        significant angular coefficients for case (a) and
        $\lfloor K/6 \rfloor$ for (b).
        The scale on the right of each plot is normalized with respect to the
        largest coefficient of the solution of highest resolution.
        As can clearly be seen in both cases, the coefficient $a_{KL}$
        converges most rapidly and does not reflect the fact that we have
        poor convergence in the radial direction.
        Since the values of the $a_{K0}$ curve in (b) are far away from the
        floating point roundoff plateau, one may question convergence and
        instead assume that at the higher radial resolutions, the very wide
        bump-type function used in \cite{rinne2016numerical} starts to get
        sampled fine enough such that the overdeterminedness of the
        equation begins to show an effect.
    }
    \label{fig:coeff_convergence_doulis_rinne}
\end{figure}

We were able to reproduce this strategy and obtained very similar results at
the resolutions mentioned in \cite{rinne2016numerical};
see figure~\ref{fig:our_doulis_rinne_q}.
Even though we can closely match the convergence of the $a_{KL}$ coefficient
(figure~\ref{fig:coeff_convergence_aNN}), we observe poor convergence
of the coefficient $a_{K0}$, i.\,e.~in the radial direction,
especially at higher resolutions (figure~\ref{fig:coeff_convergence_aKL}).%
\footnote{%
    The cause is most likely the very wide bump-type function, which
    is constructed with $b_1 = b_2 = 10^{-2}$, see equation~(3.2) in
    \cite{rinne2016numerical}.
    At low radial resolutions, the collocation points do not sample
    this function and its derivatives sufficiently fine enough,
    such that the Neumann condition is effectively ignored.
}
Furthermore, the conformal factor $\psi$ numerically determined in this
process does not pass the criterion of equation~\eqref{eq:qdn_equiv_qnd}
developed in the following subsection.
In light of these results, we were not able to confirm that the numerical
strategy presented in \cite{rinne2016numerical} successfully produces
DN-solutions.

\subsection{Our Strategy}

The basic idea here is as follows.
An ansatz of the form \eqref{eq:psi_blend} is perturbed by a
function $\chi(r,\ta)$, i.\,e.
\begin{equation}\label{eq:psi_plus_chi}
    \psi(r,\ta) = \b(r,\ta)\; \psiBL(r,\ta) + (1-\b(r,\ta))\; \psiS(r) + \chi(r,\ta),
\end{equation}
such that the DN-problem is solvable.
This latter condition is tested by {\em decomposing} the full DN-problem into
two problems, each having an existing and unique solution.
The first problem is obtained by taking \eqref{eq:poisson_q} and imposing a
Dirichlet condition just on the arches,
and a Neumann condition just on the $z$-axis for $\ta=0$ and $\ta=\pi$.
This problem is uniquely solvable
(see e.\,g.~Remark~2.1 in \cite{korotov2007two})
and we will call its solution a dn-solution $\qdn$.
The second problem is obtained by exchanging the boundary conditions and
imposing a Neumann condition on the arches and a Dirichlet condition on the
$z$-axis. Solutions of this problem will be called nd-solutions $\qnd$.
If, for some particular $\chi$ and $M$, we have
\begin{equation}\label{eq:qdn_equiv_qnd}
    \qdn \equiv \qnd,
\end{equation}
then both solutions satisfy Dirichlet and Neumann conditions on
$\partial\Omega$ and hence solve the DN-problem.
On the other hand, if \eqref{eq:qdn_equiv_qnd} is not satisfied, then
equation~\eqref{eq:poisson_q} has no DN-solution.

Naturally, the conditions to impose on $\chi$ are directly related to those of
$\psi$. Specifically, we require $\partial\chi/\partial\ta=0$ on the $z$-axis
and that $\chi$ be of bump-type in the radial direction,
i.\,e.~$\partial^n\chi/\partial r^n = 0$ for all $n \geq 0$ on the arches.

We still have to consider that the ADM mass $M$ of the glued Schwarzschild end
cannot be chosen arbitrarily---Corvino \cite{corvino2000} just proves that
there exists a mass such that the gluing can be done.
This fact manifests itself in the current construction as follows.
The DN-solution will also be a pure Neumann solution of the Poisson problem.
But for such a solution to exist, the {\em compatibility condition}
\begin{equation}\label{eq:compatibility_condition}
    \int_\Omega f \ d^3x = 0
\end{equation}
has to be satisfied, as can easily be seen when integrating
equation~\eqref{eq:poisson_q} over $\Omega$ and using Gauss's theorem on the
left-hand side.
Note that the mass $M$ enters the inhomogeneity $f$ through $\psiS$ in the
ansatz \eqref{eq:psi_plus_chi}.
We therefore read \eqref{eq:compatibility_condition} as a condition for $M$.
In \cite{rinne2016numerical}, Doulis and Rinne obtain an equivalent condition,
\begin{equation}\label{eq:integrability_condition}
    M = \int_0^\pi \int_0^\infty \left[
            \left( \frac{\partial_r \psi}{\psi} \right)^2
            + \left( \frac{\partial_\ta \psi}{r \psi} \right)^2
        \right]
        r^2 \sin\ta\ dr\ d\ta,
\end{equation}
which they call the {\em integrability condition}.
Again, equation~\eqref{eq:integrability_condition} is not an identity, since
the mass also enters the integrand through $\psi$.
As this condition is necessary for a Neumann solution to exist, there can
be no choice of $\chi$ leading to a DN-solution unless it is satisfied.

\subsection{Numerical Setup}
\label{sub:numerical_setup}

Mathematically, the task is to find a ``root'' of the map
$(\chi, M) \mapsto \qdn - \qnd$.
To guarantee the bump-condition for $\chi$, we use the ansatz
\begin{equation}\label{eq:chi_ansatz}
    \chi(r,\ta) = B_\chi(r)\, \hat\chi(r,\ta),
\end{equation}
where $B_\chi$ is a bump-type function of the form
\begin{equation}\label{eq:bump_function}
    B(r) \coloneqq \sech(s(r)),
    \qquad
    s(r) \coloneqq \frac{R}{2} \left( \frac{b_1}{r-2R} + \frac{b_2}{r-R} \right),
\end{equation}
for some $b_1, b_2 > 0$.
For $\hat\chi$ we use a truncated expansion into a product basis of Chebyshev
polynomials $T_n$ and cosines,
\begin{equation}\label{eq:chi_hat_ansatz}
    \hat\chi(r,\ta) = \sum_{i=0}^{N_r} \sum_{j=0}^{N_\ta-1}
                            a^{\chi}_{ij}\, T_i(x(r))\, \cos(2j\ta).
\end{equation}
Here, the linear map $x : [R,2R] \to [-1,1]$,
\begin{equation}\label{eq:x_of_r}
    x(r) = \frac{2}{R} r - 3,
\end{equation}
is used to transform the problem to the domain of the
Chebyshev polynomials and the cosines are used to impose
$\partial\chi/\partial\ta = 0$ on the $z$-axis.
We choose just the even cosine frequencies due to the $\pi/2$-symmetry of the
whole problem resulting from the choice of equal masses of the two
Brill-Lindquist black holes.

Using \eqref{eq:chi_hat_ansatz} to define $\hat\chi$ means that we have
$N = (N_r+1) N_\ta$ degrees of freedom, which can be used to either control
the coefficients $a^\chi_{ij}$ directly, or to specify values $\hat\chi$ should
have at, for example, the $N$ different Gauss-Lobatto collocation points
$(r(x_i), \ta_j)$, where $r(x)$ is the inverse of \eqref{eq:x_of_r} and
\begin{equation}\label{eq:chi_gauss_lobatto_points}
    \begin{stretchedarray}{LL@{\qquad}LL}
        x_i &= \cos \left( \frac{\pi i}{N_r} \right),
            & i &= 0, \ldots, N_r,
        \\
        \ta_j &= \frac{\pi j}{2 N_\ta},
            & j &= 0, \ldots, N_\ta - 1.
    \end{stretchedarray}
\end{equation}

For a particular $\chi$, we numerically compute $M$ by finding the
smallest value $M > 0$ satisfying the integrability condition
\eqref{eq:integrability_condition}.

Finally, we need to have highly accurate numerical solutions for $\qdn$ and
$\qnd$ in order to evaluate their difference. Similar to
\cite{rinne2016numerical}, we choose the pseudospectral method
\cite{boyd2001chebyshev,canuto2006spectral} and expand the
two solutions into Chebyshev polynomials for both $r$ and $\ta$
\begin{equation}\label{eq:q_trunc_expansion}
    \begin{stretchedarray}{LL}
        \qnd(r,\ta) &= \sum_{i=0}^{N^q_r} \sum_{j=0}^{N^q_\ta}
            a^{dn}_{ij}\, T_i(x(r))\, T_{2j}(y(\ta)),
        \\
        \qdn(r,\ta) &= \sum_{i=0}^{N^q_r} \sum_{j=0}^{N^q_\ta}
            a^{nd}_{ij}\, T_i(x(r))\, T_{2j}(y(\ta)),
    \end{stretchedarray}
\end{equation}
where the resolution $N^q_r, N^q_\ta$ is independent of the resolution $N_r, N_\ta$
of $\hat\chi$. Here, $y(\ta) \coloneqq 2\ta/\pi - 1$ transforms the angular
range to the domain of the Chebyshev polynomials and we have used only the
even polynomials to accommodate the $\pi/2$-symmetry.
By choosing Chebyshev polynomials for the angular ``direction'', we have the
required freedom to impose any kind of boundary condition on the $z$-axis.
For the dn- and nd-problems, all boundary conditions are imposed in the
standard way by replacing rows in the resulting pseudospectral matrix
equation.
Again, we use Gauss-Lobatto points for $x$ and also for $y$, for which they
reduce to
\begin{equation}\label{eq:cheby_lobatto_sym}
    y_j = \cos \left( \frac{\pi j}{2N^q_\ta} \right),
    \qquad
    j = 0, \ldots, N^q_\ta.
\end{equation}

\subsection{The Root Search}

The search for $\chi$ and $M$ is carried out as follows.
Let $\vc v \in \R^N$ be the vector encoding the $N$ degrees of freedom of
$\hat\chi$, e.\,g.~by specifying the values of $\hat\chi$ on the grid
\eqref{eq:chi_gauss_lobatto_points} of Gauss-Lobatto points.
This is transformed to spectral space providing the coefficients
$a^\chi_{ij}$. With $\chi$ fixed, we use the integrability condition
\eqref{eq:integrability_condition} to determine $M$. In case no such mass
exists (there are cases in which \eqref{eq:integrability_condition} cannot be
satisfied by any $M$), the search cannot continue at this point, but this
usually only occurs for too small gluing radii $R$.
With $\chi$ and $M$, we can evaluate the inhomogeneity $f$ of the Poisson
equation \eqref{eq:poisson_q} and compute the two solutions $\qdn$ and $\qnd$
with the pseudospectral method.
Sampling the difference $\qdn - \qnd$ on the Gauss-Lobatto grid of $\hat\chi$
results in $N$ values which constitute the vector $\vc w \in \R^N$.
Note that this defines a map $F : \R^N \to \R^N$, $\vc v \mapsto \vc w$.

Let $\varepsilon$ be the greater of the two absolute errors of $\qdn$ and
$\qnd$, i.\,e.
\begin{equation}\label{eq:def_epsilon}
    \varepsilon \coloneqq \max\{
        \norm{\qdn-\qdn^{\mathrm{exact}}}_\infty,
        \norm{\qnd-\qnd^{\mathrm{exact}}}_\infty
    \}.
\end{equation}
We consider $\vc v^*$ to be an approximate root of $F$ if
$\norm{F(\vc v^*)}_\infty \lesssim \varepsilon$.
Recall that the resolution $N_r, N_\ta$ of $\hat\chi$ also controls the grid
on which $\qdn - \qnd$ is measured.
If this resolution is high enough, we will have
$\norm{F(\vc v^*)}_\infty \approx \norm{\qdn-\qnd}_\infty$ and hence
\begin{equation}\label{eq:q_convergence_condition}
    \norm{\qdn-\qnd}_\infty \lesssim \varepsilon.
\end{equation}
If condition \eqref{eq:q_convergence_condition} is satisfied, we cannot
determine---at least at the chosen resolution $N^q_r, N^q_\ta$ of the
pseudospectral solutions---whether there is any difference between $\qdn$ and
$\qnd$. We will then call them identical, thus a DN-solution, within the
numerical limits. Naturally, one aims to choose the highest feasible
resolution $N^q_r, N^q_\ta$ in order to obtain a low value of $\varepsilon$.

\begin{figure}[htbp]
    \centering
    \includegraphics{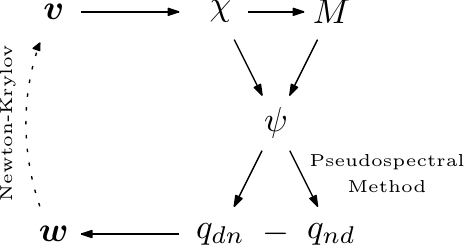}
    \caption{Diagram showing the numerical strategy for finding a DN-solution.}
    \label{fig:strategy_flow}
\end{figure}

To find a root $\vc v^*$ of $F$, we use an approximate Newton search,
the {\em Newton-Krylov} algorithm \cite{knoll2004jacobian}, which is suitable
for high-dimensional problems. Figure~\ref{fig:strategy_flow} shows the basic
scheme of the numerical strategy.
In order to enable the Newton-Krylov search to succeed, it turned out that we
had to employ means to avoid high-order contributions in the expansion of
$\hat\chi$ at the beginning of the search.
This is accomplished by starting with a low resolution for $\hat\chi$ and
applying a low-pass filter, i.\,e.~effectively damping the coefficients
$a^\chi_{ij}$ for higher $i, j$.
In successive Newton-Krylov runs,
each taking only one approximated Newton step,
the resolution of $\hat\chi$ and the dimension of the Krylov-subspace are
increased and the damping reduced until the target configuration is reached.
At this point, the Newton-Krylov algorithm takes as many steps as required to
satisfy condition~\eqref{eq:q_convergence_condition}.

Most of the code implementing this strategy has been written from scratch in
Python including the pseudospectral method.
Library routines of SciPy \cite{scipy2001}, NumPy \cite{numpy2011}, and mpmath
\cite{mpmath2013} were used for the root finding, integration, and matrix
equation solving.

\subsection{Testing the Convergence on an Exact Solution}

A crucial aspect of the above strategy is to obtain both a high accuracy and
a good estimation of the error $\varepsilon$ of the pseudospectral solutions
$\qdn$ and $\qnd$.
The first of these goals is achieved by choosing a sufficiently high resolution
$N^q_r, N^q_\ta$, while for the second goal we employ a detailed convergence
analysis and compare the solutions with a known exact solution of a very
similar problem.
Such a test problem is provided by the analytical solution for an
approximated inhomogeneity $f$ obtained in \cite{giulini2005corvino}.
For the case $M = 2m$, an explicit solution formula can be given.
To this end, consider the ansatz \eqref{eq:psi_blend} for $\psi$ and choose
$\b$ as
\begin{equation}\label{eq:beta_ansatz}
    \b(r,\ta) = \a(r) + \mu(r) \sin^2\ta,
\end{equation}
where $\a$ satisfies \eqref{eq:beta_conditions_a} and \eqref{eq:beta_conditions_b}
and $\mu$ is of bump-type.
Expanding now the inhomogeneity in the inverse gluing radius $R$ up to first
order, Giulini and Holzegel obtain an ordinary differential equation for $\mu$
and conditions for $\a$. For $M = 2m$, the result can be written as
(cf.~\cite{pookkolb2018numerical})
\begin{subeqns}[\label{eq:alpha_mu_solution}]
    \begin{eqnarray}
        \label{eq:alpha_conditions}
        \int_R^{2R} t^{-(2-\sqrt3)} \a'(t)\ dt = 0,
        \qquad
        \int_R^{2R} t^{-(2+\sqrt3)} \a'(t)\ dt = 0,
        \\
        \begin{stretchedarray}[2.2]{LL}
            \label{eq:mu_solution}
            \mu(r) =& - \frac{1}{2} r^2 \a''(r) + \frac{3}{2} r \a'(r) \\
                      &+ \int_R^r \a'(t) \left(
                            \frac{3+2\sqrt3}{2}
                                \left( \frac{r}{t} \right)^{2-\sqrt3}
                            + \frac{3-2\sqrt3}{2}
                                \left( \frac{r}{t} \right)^{2+\sqrt3}
                        \right)\ dt.
        \end{stretchedarray}
    \end{eqnarray}
\end{subeqns}
To satisfy the conditions \eqref{eq:alpha_conditions} for $\a$ we choose
\begin{equation}\label{eq:alpha_GH}
    \a(r) = \a_0(r) + B_\a(r) (c_0 + c_1 r),
\end{equation}
where $B_\a$ is a bump-type function of the form \eqref{eq:bump_function} and
$\a_0$ is given by
\begin{equation}\label{eq:alpha0}
    \a_0(r) \coloneqq \frac{1}{2}\big( 1 + \tanh(s(r)) \big).
\end{equation}
The constants $c_0, c_1$ are found via a simple Newton search for a root of
the map $(c_0, c_1) \mapsto (I_1, I_2)$, where $I_1, I_2$ are the numerical
values of the integrals in \eqref{eq:alpha_conditions}.
Using the results of \cite{giulini2005corvino}, one can then easily write down
the exact solution $\tilde q$ of the approximated problem in terms of $\a$ and
$\mu$.
Note that $\tilde q$ is a DN-solution and that we therefore expect
the pseudospectral solutions $\tilde\qdn$ and $\tilde\qnd$ of this
approximated problem to coincide.

\begin{figure}[tbp]
    \centering
    \subfigure[]{
        \includegraphics[trim=2 10 10 0,clip,width=.5\textwidth]{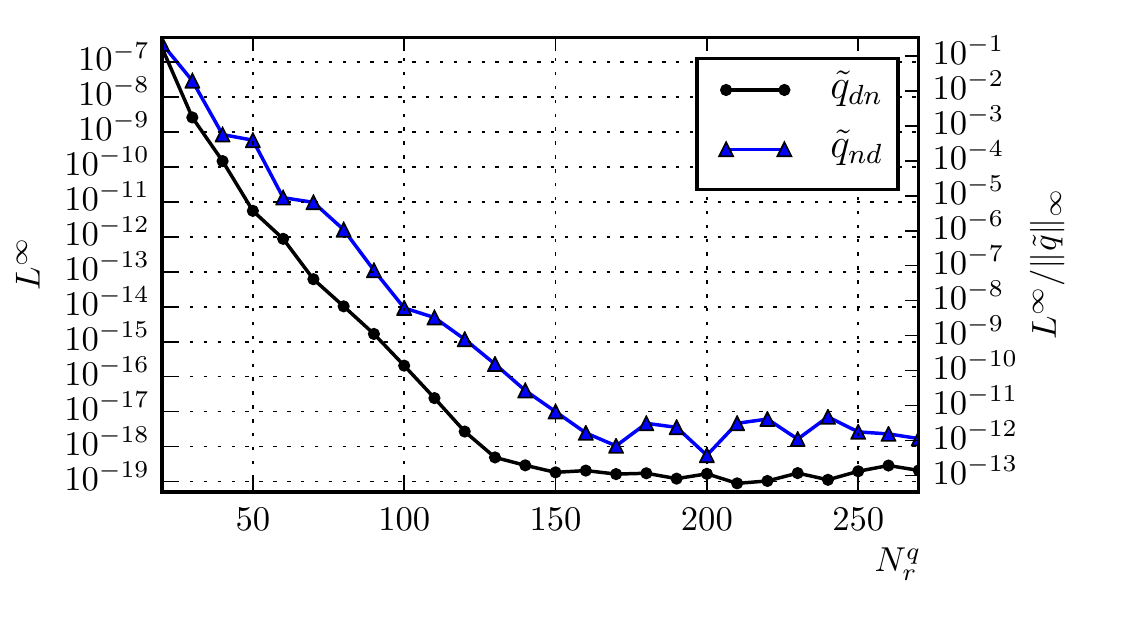}
        \label{fig:q_tilde_convergence_a}
    }%
    \subfigure[]{
        \includegraphics[trim=0 10 14 0,clip,width=.5\textwidth]{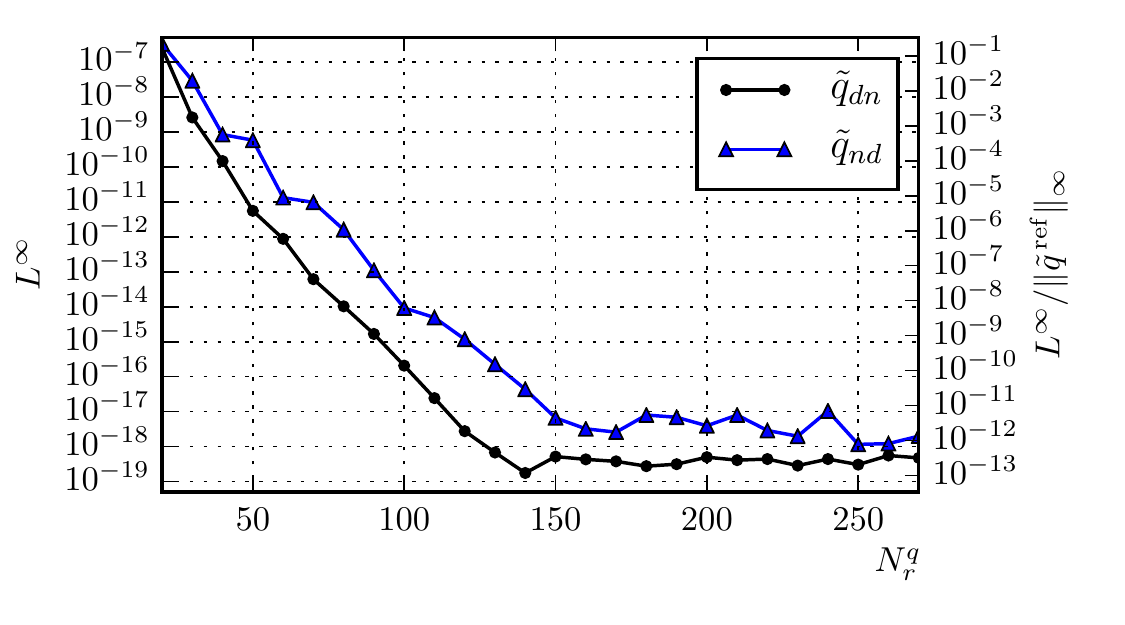}
        \label{fig:q_tilde_convergence_b}
    }
    \caption{%
        Convergence of the pseudospectral solutions $\tilde\qdn$ and
        $\tilde\qnd$ to the exact solution $\tilde q$ in (a) and to the
        solutions $\tilde\qdn^{\,\mathrm{ref}}$ and
        $\tilde\qnd^{\,\mathrm{ref}}$, respectively, in (b).
        The reference solutions for (b) were computed with a resolution of
        $N^q_r = 280$.
        For all solutions, we chose $N^q_\ta$ to be the integer part of
        $N^q_r/6$.
        The following parameters were used: $d=10$, $m=2$, $M=2m=4$,
        $R=5000$, and $b_1 = b_2 = 2$ for both $\a_0$ and $B_\a$.
    }
    \label{fig:q_tilde_convergence}
\end{figure}

Figure~\ref{fig:q_tilde_convergence} shows that this is in fact the case up
to a remaining roundoff error of about $10^{-17}$ and that the convergence for
increasing resolution is exponential as would be expected.
It also exhibits the typical roundoff plateau which occurs when the floating
point errors dominate.
Note that figure~\ref{fig:q_tilde_convergence_b} provides a very good estimate
of the accuracy despite being a purely intrinsic test without knowledge of the
exact solution $\tilde q$ of the approximated DN-problem.

\section{Numerical Results}

\begin{figure}[tbp]
    \hfill
    \subfigure[]{
        \includegraphics[trim=50 18 15 20,clip,width=.4\textwidth]{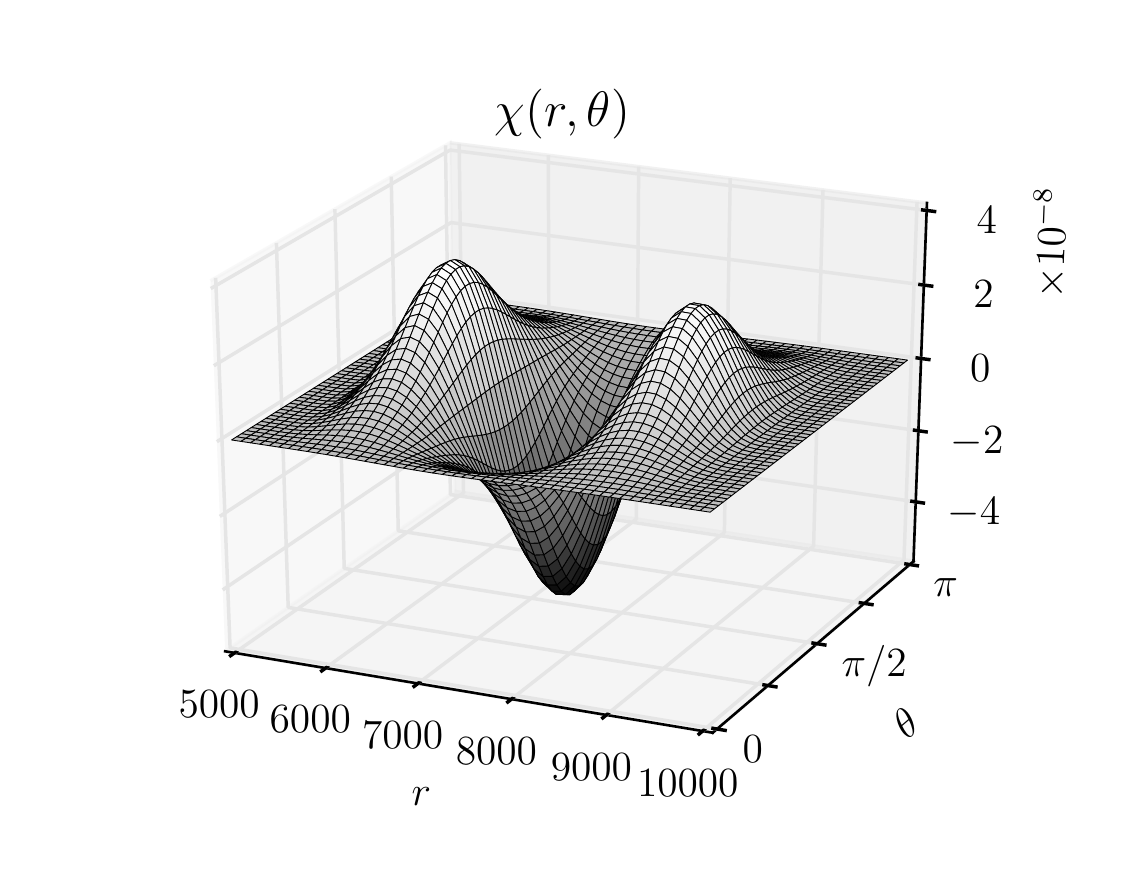}
        \label{fig:DNsolution1_chi}
    }
    \hfill
    \subfigure[]{
        \includegraphics[trim=50 18 15 20,clip,width=.4\textwidth]{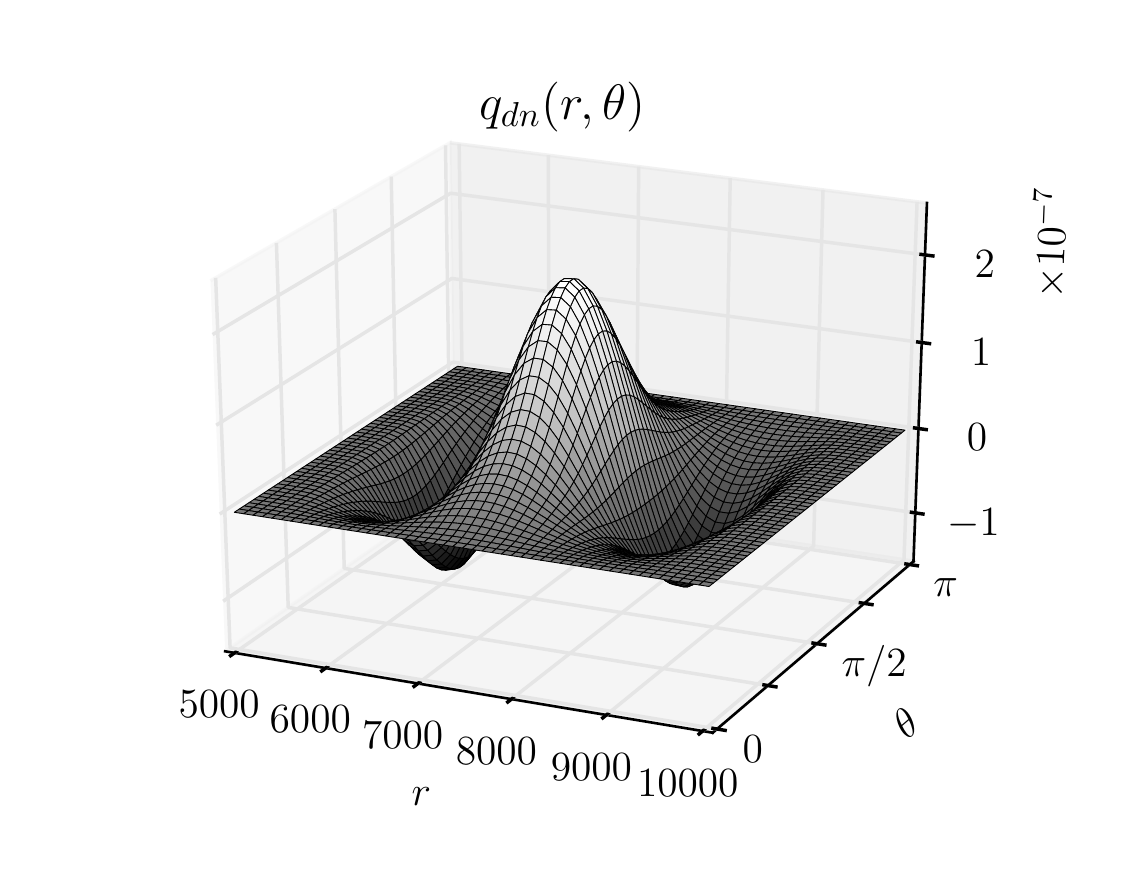}
        \label{fig:DNsolution1_q}
    }
    \hfill
    \caption{%
        Solution of the numerical search for $\chi$ depicted in (a).
        In (b), the dn-solution is shown which is equal to the nd-solution
        within the numerical limits.
    }
    \label{fig:DNsolution1}
\end{figure}

\begin{figure}[tbp]
    \centering
    \includegraphics[trim=2 10 10 0,clip,width=0.5\linewidth]{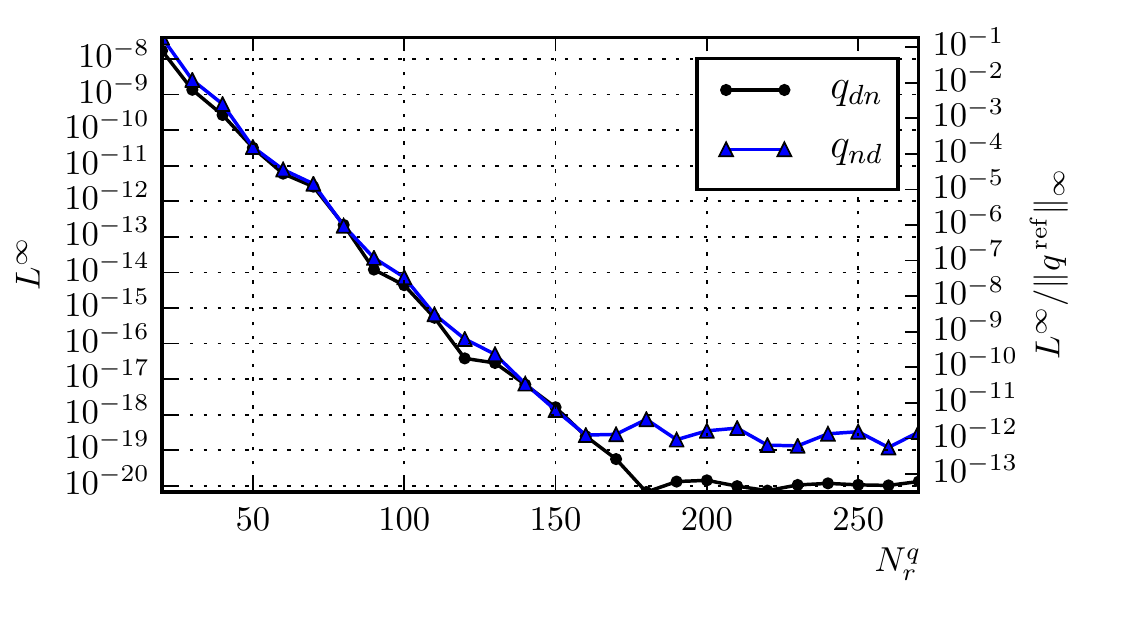}
    \caption{%
        Convergence of the dn- and nd-solutions to the reference solutions
        $\qdn^{\,\mathrm{ref}}$ and $\qnd^{\,\mathrm{ref}}$, respectively,
        computed with $N^q_r = 280$, $N^q_\ta = 46$.
    }
    \label{fig:DNsolution_convergence}
\end{figure}

Here we present numerical results for the non-approximated
DN-problem.
Starting the search for $\chi$ and $M$ with the ansatz \eqref{eq:psi_plus_chi}
for $\psi$ and choosing $\b(r,\ta) = \a_0(r)$ as in \eqref{eq:alpha0} with
$b_1 = b_2 = 2$ and the configuration $d=10$, $m=2$, and $R=5000$,
we obtain a solution for the deformation $\chi$ depicted in
figure~\ref{fig:DNsolution1_chi}.
During the search, the resolution for the pseudospectral solutions has been
chosen to be $N^q_r = 160$, $N^q_\ta = 26$.
Figure~\ref{fig:DNsolution1_q} shows the dn-solution $\qdn$, which is close to
the nd-solution $\qnd$ with
$\norm{\qdn - \qnd}_\infty \approx 2.4 \times 10^{-19}$.
This is compatible with the accuracy read off from
figure~\ref{fig:DNsolution_convergence} such that condition
\eqref{eq:q_convergence_condition} is satisfied and therefore
provides a strong indication that this solution is close to an exact
solution of the DN-problem.
The ADM mass $M$ of the glued Schwarzschild end in this result reads
\begin{equation}\label{eq:DNsolution1_mass}
    M = 2m + (7.5620 \pm 0.0001) \times 10^{-10}.
\end{equation}

Using the same interior configuration $m$, $d$, and gluing radius $R$,
but different search parameters (initial resolution for $\hat\chi$, etc.),
we were able to find different combinations of deformations $\chi$ and ADM
masses $M$ which still satisfy the condition \eqref{eq:q_convergence_condition}.
The masses found thus far range from about
$2m + 1.8 \times 10^{-11}$
to
$2m + 4.7 \times 10^{-8}$.

\section{Conclusion}
The above results provide strong indication that the presented numerical
strategy is able to successfully produce a highly accurate gluing of a
Brill-Lindquist interior to the exterior  Schwarzschild metric so as
to everywhere satisfy the time-symmetric vacuum constraints. Moreover,
the way this strategy is set up leaves a large freedom for how the final
gluing is done. As this freedom is a genuine feature of Corvino's
construction, it should clearly be preserved, at least to some extent,
in any more or less faithful numerical implementation, like the one 
presented here. This freedom then opens up the possibility to impose 
further conditions on the gluing, depending on the desired properties 
one wishes the final metric to share. One such property, that we already 
used as motivation in the introduction, is the minimisation of the 
ADM mass of the glued Schwarzschild end. Such a minimisation is taken 
to signal the removal of spurious (``junk'') gravitational radiation.
On the other hand, it seems clear that just picking \emph{some} gluing 
function will generically add rather than subtract gravitational waves 
and hence result in overall masses $M>2m$, as we have seen above and 
as was also seen in \cite{rinne2016numerical}. But, as shown in our paper,
numerical implementations exist which preserve the freedom of
Corvino's construction to a large extent and which may eventually be
used to adjust the gluing so as to actually reduce the overall mass
$M$ below $2m$.
Presently we do not know how to effectively translate
such an energy-minimisation condition into the numerical strategy
in a systematic (i.e. not just based on trial and error) way, but our work 
suggests that this should be possible.

Finally we mention that quite independently of the question of mass
reduction,  Corvino-type gluing constructions may quite generally find
interesting applications in numerical evolution codes. Being exactly
Schwarzschild in a neighbourhood of spatial infinity, smooth null infinities
$\Scri^+$ are guaranteed.  This allows evolutions of hyperboloidal
slices extending to $\Scri^+$, which in turn enable unambiguous
extractions of gravitational radiation data. A detailed description of
such a scheme is provided by Doulis and Rinne in \cite{rinne2016numerical}.
\section*{References}


\end{document}